\documentclass{PoS}

\usepackage{amsmath}
\usepackage{lineno}

\title{Monitoring the TeV sky on hours long timescales with HAWC}

\ShortTitle{Monitoring the TeV sky on hours long timescales with HAWC}

\author{\speaker{Israel Martinez-Castellanos}$^a$ for the HAWC Collaboration$^b$\\
        $^a$University of Maryland\\
        $^b$For a complete author list see \href{www.hawc-observatory.org/collaboration/icrc2017.php}{www.hawc-observatory.org/collaboration/icrc2017.php}\\
        E-mail: \email{imc@umd.edu}}

\abstract{The High-Altitude Water Cherenkov Observatory (HAWC) is a large field of view ($\sim$2sr) instrument sensitive to very-high energy gamma rays ($\sim$0.5-100TeV). It is located in central Mexico (19$^\circ$N) and has a high duty cycle ($\sim$95\%). These characteristics allow it to continuously monitor $\sim$2/3 of the sky, looking for transient events, such as flares from Active Galactic Nuclei or possibly other unknown phenomena. Presented here is an unbiased real-time monitoring on hours timescales which provides daily flux measurements for all locations in our observable sky promptly after they leave our field of view. These measurements are then used to follow known TeV sources and to perform a blind search. The alerts generated from these analyses, some of which have been made public through the Astronomer's Telegraph, can trigger small field of view instruments, enabling deep observations of sources during their high state activity which can constrain acceleration mechanisms. Additionally, these measurements allow us to respond quickly to external alerts.}

\FullConference{35th International Cosmic Ray Conference --- ICRC2017\\
		10--20 July, 2017\\
		Bexco, Busan, Korea}

\begin{document}

\section{Introduction}

The TeV gamma-ray sky is known to contain variable sources, mostly binaries \cite{HESSJ0632+057}\cite{3p9modulation} and active galactic nuclei (AGN) \cite{exceptionalFlare}\cite{magic501}, with variability timescales ranging from minutes to months. AGN variability can be particularly extreme with flaring states reaching fluxes equivalent to several times the Crab Nebula. Most of these observations have been made with Imaging Atmospheric Cherenkov Telescopes (IACTs) that, although being the most sensitive very-high energy (VHE) gamma-ray instruments, have a very narrow field of view and  therefore provide measurements biased towards flares detected in other wavelengths.

HAWC is a wide field of view gamma-ray detector that covers $\sim2/3$ of the sky every day with $>95\%$ duty cycle, allowing to look for variability in the TeV gamma-ray sky in an unbiased manner. It is sensitive to gamma rays from  $\sim$500GeV to $\sim$100TeV, with an angular resolution ranging from $\sim0.2^\circ$ to $\sim1^\circ$, depending on the quality of the event. HAWC is located at a latitude of 19$^\circ$N, in Sierra Negra, M\'exico, at an altitude of 4,100 meters. More details on the experiment, event reconstruction, performance and systematics can be found in Abeysekara, et al. \cite{crabPaper}. 

Here we present an analysis performed in all the observable sky on a daily basis. It has been running online since late 2016, with the goal of informing other experiments in the field about transient TeV emission, enabling deep observations that might otherwise not be possible. 

\section{Search method}

We present here a real-time search for hours long variability from all points in the HAWC field of view. This search runs in parallel with HAWC's existing real-time searches for short timescale flares from known TeV and 2FHL sources \cite{flareMonitorPaper} as well as the all-sky search for rapid bursts on seconds timescales \cite{grbProceeding}. The search is performed by integrating each point on the sky over a full transit, defined as the time spent at $<45^\circ$ from zenith, and testing for an event excess relative to the steady cosmic ray background. A location in the sky that transits directly overhead is exposed for 6.3hr, decreasing to 5hr for a point that culminates at 35$^\circ$ from zenith, making this effectively a search corresponding to a few hours timescale. 

The algorithm used to compute the excess at each point on the sky is the same as \cite{crabPaper} but with an optimized code fast enough to run in near real-time. It works by generating maps of detected events every 15 minutes with a background estimate given by the method described in Section 3.1. Every location that sets during this time period is then analyzed according to the method in Section 3.2. This process takes <5min, so the results are available for each location 5 to 20 min after they leave our field of view, depending on their right ascension (RA). 

During the analysis we divide our data into 9 analysis bins of increasing fraction of PMTs hit ($fhit$)\footnote{A mentioned in \cite{crabPaper} the lowest $fhit$ bin currently starts at 6.7\% of the detector, or $\sim$70 PMTs. There are planned developments what would allow us to use events with 20-30 PMTs hit, which would improve HAWC sensitivity in the $\sim$100-500 GeV range.}. As $fhit$ increases the cosmic-ray rejection capability improves rapidly, causing the all-sky rate of events passing the gamma-ray discrimination cuts to drop dramatically, from $\sim$500Hz to $\sim$0.05 Hz. This sets some of the analysis bins well in the Poisson regime when integrating for a few hours, a fact that will become relevant in the following sections. See \cite{crabPaper} for definition and characterization details of each analysis bin.

\subsection{Background calculation}
\label{SectionBkg}

Before the map-making step, a set of cuts are applied to the data in order to discriminate between gamma rays and cosmic rays --see \cite{crabPaper} for the method description and performance--. Nevertheless, cosmic rays still dominate our background. We estimate the background (bkg) using a method called \textit{direct integration} \cite{milagroDirectIntegration}, which consists on convolving in time ($t$) the total event rate with the efficiency (eff) of the detector in local coordinates 

\begin{align}
\mathrm{bkg}(\mathrm{RA},\mathrm{Dec}) 
&= \int \mathrm{rate}(t) \mathrm{eff}(\mathrm{HA},\mathrm{Dec}) dt \nonumber \\
&= \int \mathrm{rate}(t) \mathrm{eff}(t-\mathrm{RA},\mathrm{Dec}) dt ,
\end{align}

where RA stands for right ascension, Dec for declination and HA for the local hour angle. The efficiency of the detector, also known as acceptance, is the probability that an event came from a given location, in local coordinates, so it is normalized to 1. This method assumes that this function is independent of the total rate, and varies slowly over time. 

For the analysis bins with low $fhit$ we estimate the efficiency with a histogram in (HA,Dec) using a 2hr integration period, as done originally in \cite{milagroDirectIntegration}. This provides enough statistics while accounting for the cosmic ray anisotropy, giving the precision required for these analysis bins with low signal to noise ratio. For analysis bins with high $fhit$ the number of events observed is so scarce that we need to integrate for longer periods of time and smooth the resulting efficiency maps. Different integration times and smoothing schemes have been applied on this low statistics analysis bins without affecting the result significantly.

\subsection{Flux and significance calculation}
\label{SectionSig}

For each analysis bin, the data is mapped into a HEALPix structure \cite{healpixPaper} with $N_{side} = 1024$ ($\sim0.06^\circ$ radius pixels). We then apply a binned maximum likelihood ratio analysis, testing the hypothesis of a source located at the center of each pixel versus a background only null hypothesis. The likelihood function is formed based on a Poisson distribution

\begin{align}
\mathcal{L}(f) = \sum_{i=1}^N \frac{(b_i+e_i f_0)^{d_i} \exp\left( - (b_i + e_i f_0) \right)}{d_i!} ,
\end{align}

where

\begin{itemize}
\setlength\itemsep{0em}
\item $f_0$ is the flux normalization factor
\item the index $i$ runs over all analysis bins and pixels near the tested location\footnote{We first try with a padding of $1^\circ$, and if $TS>9$, we extend it to $3^\circ$. This saves computational resources while accounting for the tails of the point spread function for strong signals, in which case they do matter.} \\--i.e. $N = \#analysis~bins \times \#pixels$--
\item $b_i$ are the estimated background counts
\item $e_i$ are the normalized expected excess counts.
\item $d_i$ are the number of counts observed
\end{itemize}

\begin{figure}
\centering
\includegraphics[width=0.7\textwidth]{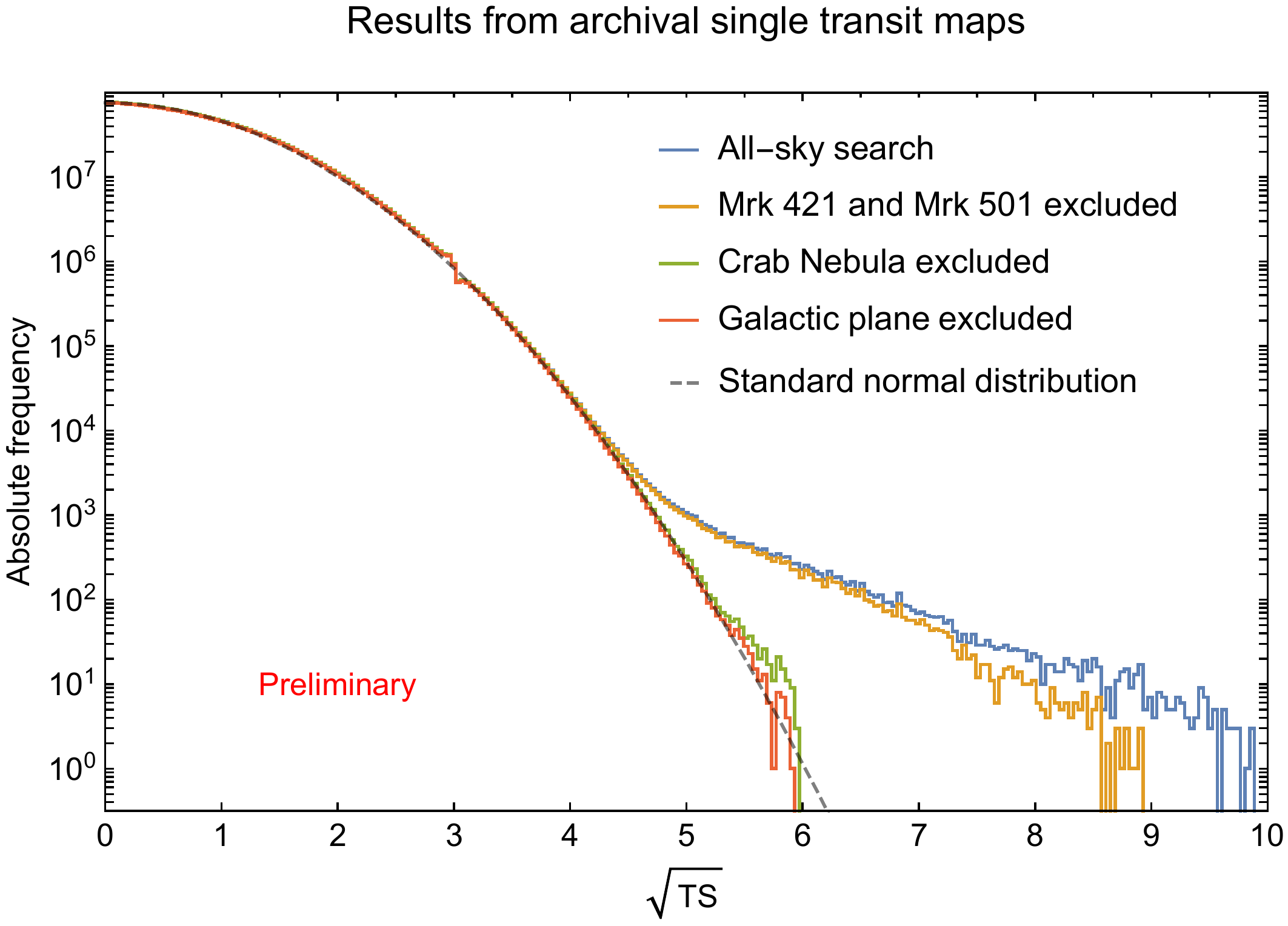}
\caption{$\sqrt{TS}$ distribution for 765 single transit maps from archival data, covering from November 2014 to January 2017. A series of regions of exclusion are applied progressively to show where the excess is coming from. The small indents at $\sqrt{TS} = 3$ is an artifact of the change of padding that occurs when $TS = 9$.}
\label{Figure:sigHist}
\end{figure}

The expected number of excess counts is estimated from Monte Carlo simulations. A simple power law -- i.e. $f(E) = f_0 (E/E_0)^{-\alpha}$, where $f_0$ is the flux at energy $E_0$ and $\alpha$ is the spectral index-- is used as the spectral hypothesis. We then form a TS by fixing the spectral index and fitting the flux norm

\begin{align}
TS = 2 \ln \frac{\max \mathcal{L}(f_0)}{\mathcal{L}(f_0=0)} = 2 \ln \frac{\mathcal{L}(F_0)}{\mathcal{L}(f_0=0)}
\end{align}

where $F_0$ denotes the best estimate for the flux normalization. We restrict the computation to positive values of $F_0$, and therefore set $TS=0$ for under-fluctuations. $TS$, according to Wilk's theorem, is approximately distributed as a $\chi^2$ distribution with one degree of freedom, in which case $\sqrt{TS}$ has a standard normal distribution. Fig.\ref{Figure:sigHist} shows that this is a good description in our case, with a caveat: $\sim$60\% of the time the result is $TS=0$, as opposed of the 50\% we might expect. This happens because Poissonian statistics come into play for this short timescales, something that disappears when we integrate years of data \cite{catalogPaper}. Nevertheless, the effect is negligible for large $TS$ values --e.g. $\sqrt{TS} = 4.96$ is equivalent to $5\sigma$ based on its p-value--.

\section{Sensitivity}

\begin{figure}
\centering
\includegraphics[width=0.7\textwidth]{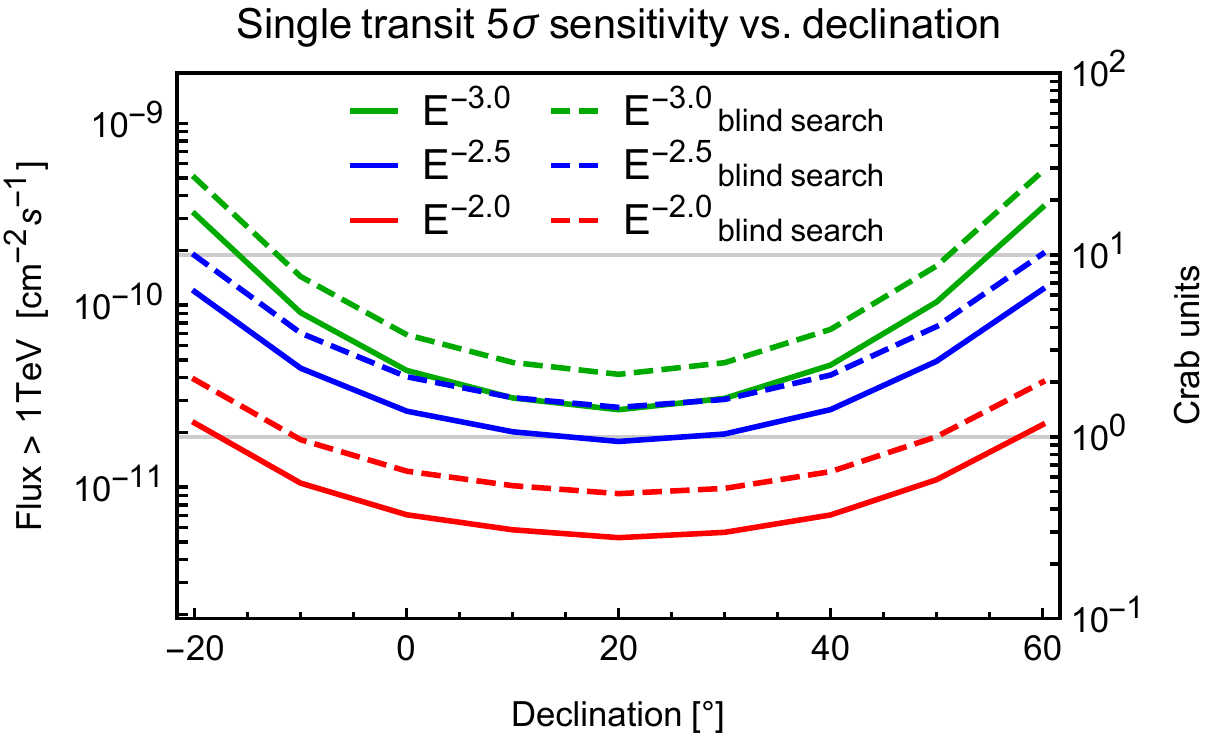}
\caption{Sensitivity of the single transit search, defined as the flux that have a median significance of $5\sigma$, for various spectra. The solid lines assume a known location, while the dashed line corresponds to $7.6\sigma$ to account for the trials involved in a blind search, as discussed in the text, except temporal trials.}
\label{Figure:sensitivity}
\end{figure}

We characterize the sensitivity of this search by quoting the flux that would result in 5$\sigma$ half of the time. We estimated this using simulations and the results are shown in Fig. \ref{Figure:sensitivity}.

Since this is a blind search, we need to take into account the trials involved when computing the sensitivity. The biggest source of trials is the unknown localization of the event. We analyze $\sim8.3\times10^6$ positions in the sky. Although simulation studies have shown that the effective number of independent locations is around an order of magnitude lower, if we are conservative and take each pixels as an independent trial we would need $7.5\sigma$ pre-trials for a $5\sigma$ post-trial detection. 

Another, albeit smaller source of trials is the multiple spectral index hypotheses that are tried. Instead of fitting the spectral index, which would slow the calculation, we try three different hypotheses, $\alpha = $-2, -2.5 and -3, and select the one with the best TS. Although the results are highly correlated, we conservatively take them as independent trials, which result in a needed pre-trail significance of $7.6\sigma$. 

Lastly, before claiming a detection, we would need to compute the temporal trials incurred depending on the time we have been looking for, approximately equal to the number of days times the search duration. We currently look for 1, 2 and 3 transits combined. It is however worth noticing that one of the motivations for this search is to alert other experiments, so sub-threshold events with a low false alarm rate can be made public in order to motivate deep observations.  
 
\section{Results}

So far, all the detections by the all-sky search are consistent with coming from the Crab Nebula, Markarian 421 and Markarian 501, all strong sources known to emit in TeV, and reported in the latest HAWC catalog \cite{catalogPaper}. The Crab is detected on average at $\sim5\sigma$ every day and it is consistent with a constant flux; on the other hand, the quiescent flux of Mrk 421 and Mrk 501 is much smaller, but they show strong variability, reaching multiple times the Crab Nebula flux \cite{variabilityPaper}.

\begin{figure}
\centering
\includegraphics[width=0.7\textwidth]{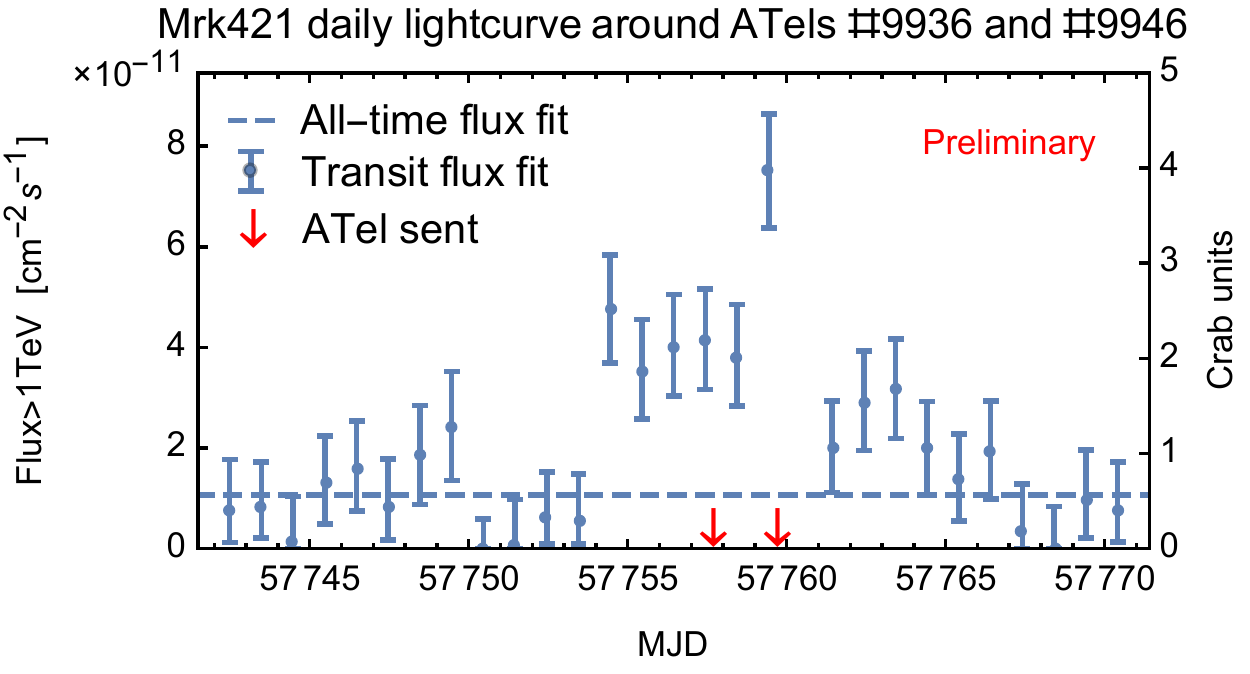}
\caption{Mrk 421 daily light curve from December 12th, 2016, to January 17th, 2017. Error bars correspond to 1$\sigma$ Feldman-Cousins confidence intervals.}
\label{Figure:lightcurve}
\end{figure}

As shown in Fig.\ref{Figure:sigHist}, they represent the bulk of the excess with respect to the null hypothesis in the one transit search. Because of this, they are monitored individually using the same set of maps produced for the all-sky search. We've alerted the community of enhanced detected emission from Mrk 421 and Mrk 501 trough the Astronomer's Telegraph\footnote{ATels \href{http://www.astronomerstelegram.org/?read=8922}{\#8922}, \href{http://www.astronomerstelegram.org/?read=9137}{\#9137},  \href{http://www.astronomerstelegram.org/?read=9936}{\#9936} and \href{http://www.astronomerstelegram.org/?read=9946}{\#9946}}. A light curve corresponding to the latest alerts is shown in Fig. \ref{Figure:lightcurve}.

In addition to these self-triggered alerts, the maps produced online are available to follow-up external alerts. In particular, they are used to search for counterparts of very-high energy neutrinos and gravitational waves. For the latter, it is a search complementary to the main search performed in the seconds timescale \cite{}. 

\section{Conclusion and outlook}

We presented the current analysis performed on all locations inside HAWC field of view everyday, corresponding to a typical integration time in the order of a few hours. We've shown that it has a $5\sigma$ sensitivity for a flux equivalent of that of the Crab Nebula, varying depending on the source spectrum and declination.

So far the events detected with this search have been consistent with known sources. When they represented an enhanced emission they have been communicated to the community though the Astronomer's Telegraph. In the future HAWC will continue to alert the community on this occasions, and possibly about other transient TeV emitters currently unknown, with the goal of triggering deep observations by other experiments in the field.

Current efforts are focused on sharing sub-threshold events with partner experiments, in order to achieve the quick reaction times these follow-up observations might require. Similar searches in other times scales, from one hour to a month, are currently under development as well.

\section{Acknowledgments}

We	acknowledge	the	support	from:	the	US	National	Science	Foundation	(NSF);	the	
US	Department	of	Energy	Office	of	High-Energy	Physics;	the	Laboratory	Directed	
Research	and	Development	(LDRD)	program	of	Los	Alamos	National	Laboratory;	
Consejo	Nacional	de	Ciencia	y	Tecnolog\'{\i}a	(CONACyT),	M{\'e}xico	(grants	
271051,	232656,	260378,	179588,	239762,	254964,	271737,	258865,	243290,	
132197),	Laboratorio	Nacional	HAWC	de	rayos	gamma;	L'OREAL	Fellowship	for	
Women	in	Science	2014;	Red	HAWC,	M{\'e}xico;	DGAPA-UNAM	(grants	RG100414,	
IN111315,	IN111716-3,	IA102715,	109916,	IA102917);	VIEP-BUAP;	PIFI	2012,	
2013,	PROFOCIE	2014,	2015; the	University	of	Wisconsin	Alumni	Research	
Foundation;	the	Institute	of	Geophysics,	Planetary	Physics,	and	Signatures	at	Los	
Alamos	National	Laboratory;	Polish	Science	Centre	grant	DEC-2014/13/B/ST9/945;	
Coordinaci{\'o}n	de	la	Investigaci{\'o}n	Cient\'{\i}fica	de	la	Universidad	
Michoacana. Thanks	to	Luciano	D\'{\i}az	and	Eduardo	Murrieta	for	technical	
support.

\end{document}